\documentclass[twocolumn,twoside]{Jornadas}
\usepackage[utf8]{inputenc}
\usepackage[english]{babel}
\usepackage{graphicx}
\usepackage{amsmath}
\usepackage{color}
\usepackage{caption}
\usepackage[caption=false,font=footnotesize]{subfig}
\usepackage{placeins}
\usepackage{hyperref}
\usepackage{url}
\usepackage{booktabs}

\graphicspath{{.}{./Figuras/}}

\begin{document}

\makeatletter
\def\abstract{%
    \if@twocolumn
      \small\it Abstract\/\bf---$\!$%
    \else
      \begin{center}\vspace{-0.8em}\small\bf Abstract\end{center}\quotation\small
    \fi}
\def\keywords{\vspace{-.3em}
    \if@twocolumn
      \small\it Keywords\/\bf---$\!$%
    \else
      \begin{center}\small\bf Keywords\end{center}\quotation\small
    \fi}
\def\thebibliography#1{\section*{References}\footnotesize\list
    {[\arabic{enumi}]}{\settowidth\labelwidth{[#1]}\leftmargin\labelwidth
    \advance\leftmargin\labelsep \itemsep 0pt plus .5pt
    \usecounter{enumi}}
    \def\newblock{\hskip .11em plus .33em minus .07em}
    \sloppy\clubpenalty4000\widowpenalty4000
    \sfcode`\.=1000\relax}
\def\fnum@table{Table~\thetable}
\makeatother

\title{FRAGATA: Semantic Retrieval of HPC Support Tickets via Hybrid RAG over 20 Years of Request Tracker History}

\author{%
     Santiago Param\'{e}s-Est\'{e}vez\thanks{A Spanish version of this paper has been accepted at Jornadas SARTECO 2026. Code is available at: \href{https://github.com/s-parames/fragata}{https://github.com/s-parames/fragata}}\thanks{Galicia Supercomputing Center (CESGA), Santiago de Compostela, e-mail: {\tt sparames@cesga.es}.},
     Nicol\'{a}s Filloy-Montesino\thanks{Universidade de Vigo.},
     Jorge Fern\'{a}ndez-Fabeiro$^2$
     and Jos\'{e} Carlos-Mouri\~{n}o Gallego$^2$%
}

\maketitle
\markboth{}{}
\pagestyle{empty}
\thispagestyle{empty}

\begin{figure}[t]
\centering
\includegraphics[width=0.55\columnwidth]{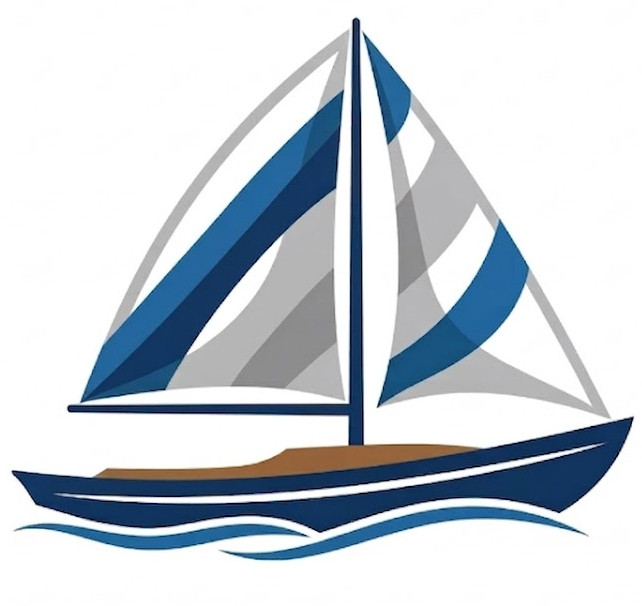}
\end{figure}

\begin{abstract}
The technical support team of a supercomputing centre accumulates, over the course of decades, a large volume of resolved incidents that constitute critical operational knowledge. At the Galicia Supercomputing Center (CESGA) this history has been managed for over twenty years with Request Tracker~(RT), whose built-in search engine has significant limitations that hinder knowledge reuse by the support staff. This paper presents \textsc{Fragata}, a semantic ticket search system that combines modern information retrieval techniques with the full RT history. The system can find relevant past incidents regardless of language, the presence of typos, or the specific wording of the query. The architecture is deployed on CESGA's infrastructure, supports incremental updates without service interruption, and offloads the most expensive stages to the FinisTerrae~III supercomputer. Preliminary results show a substantial qualitative improvement over RT's native search.
\end{abstract}


\section{Introduction}

High-Performance Computing (HPC) centres operate heterogeneous platforms hosting hundreds of scientific applications, compilers, parallel libraries, resource schedulers, and distributed storage systems. In this environment, the technical support team acts as a critical interface between research users and the infrastructure, resolving incidents ranging from MPI compilation errors to storage quotas, Conda environment configuration, or network failures on GPU nodes. The Galicia Supercomputing Center (CESGA) has managed this workflow for over two decades using Request Tracker (RT)~\cite{rt}, a widely used open-source ticketing tool.

RT offers two main interaction channels: a web interface (available only to support staff) and e-mail. CESGA users interact with the system exclusively via e-mail, sending and replying to messages that RT automatically converts into tickets and associated responses. This dynamic has direct implications for the quality of the stored text, as detailed in Section~\ref{sec:normalization}.

The history accumulated by CESGA in RT spans over twenty years of conversations between support staff and users, constituting an invaluable operational memory: a significant proportion of the incidents received today are variants of problems solved years ago. However, RT version 4.4.1 has severe limitations in its built-in search: it does not index the full ticket body, is case-sensitive, does not tolerate typos, does not normalize morphological variants, and lacks any notion of semantic similarity. As a result, both veteran and newly onboarded staff struggle to locate relevant past cases, leading to duplicated effort, loss of institutional knowledge, and increased mean resolution time.

This paper presents \textsc{Fragata}, a semantic ticket retrieval system designed and deployed at CESGA to mitigate these limitations. \textsc{Fragata} applies the \emph{Retrieval-Augmented Generation} (RAG) paradigm~\cite{lewis2020rag}, an approach that improves search quality by first retrieving the most relevant documents from a knowledge base and presenting them as context. Specifically, the system combines dense retrieval based on \emph{embeddings}, numerical vector representations of text meaning, with classical BM25 lexical retrieval~\cite{robertson2009bm25} and \emph{reranking} via \emph{cross-encoders}~\cite{nogueira2019passage}. The system also integrates complementary sources (the centre's technical documentation, scientific application manuals, and repositories) and is deployed on a hybrid architecture comprising a virtual machine and the FinisTerrae~III supercomputer, with incremental ingestion and hot-reload of the search engine.

The main contributions are: (i)~the design of a reproducible protocol for extraction, normalization, and chunking of RT's SQL history; (ii)~a hybrid retrieval architecture with weighted fusion and query-aware reranking; (iii)~a hot-swap mechanism that guarantees continuous availability during re-indexing; and (iv)~operational integration with an HPC scheduler to offload expensive ingestion stages without penalizing service latency.

\section{Related work}

\subsection{Neural information retrieval}

Information retrieval has undergone a profound transformation in recent years thanks to \emph{transformer}-based models~\cite{vaswani2017attention}. These neural network architectures, which underpin models such as BERT~\cite{devlin2019bert}, learn to represent texts as dense numerical vectors (\emph{embeddings}) that capture the semantic meaning of words and sentences. Two texts with similar meaning yield vectors that are close in the embedding space, enabling the retrieval of relevant documents by measuring vector distances rather than relying on exact word matches. In particular, Sentence-BERT~\cite{reimers2019sentencebert} adapted BERT to efficiently produce sentence-level embeddings, and dense retrieval systems such as DPR~\cite{karpukhin2020dpr} demonstrated that this approach can outperform traditional lexical search on question-answering tasks.

However, purely dense retrieval exhibits weaknesses on queries containing highly specific terminology, proper names, or technical identifiers, where exact word matching remains decisive. Consequently, hybrid approaches that combine BM25, a classical retrieval algorithm based on term frequencies, with dense retrieval~\cite{lin2021pyserini} have become the practical state of the art. These systems benefit especially from a subsequent \emph{reranking} stage, in which a \emph{cross-encoder}~\cite{nogueira2019passage} jointly evaluates the query and each candidate to reorder the results with greater precision.

The RAG paradigm~\cite{lewis2020rag,gao2024ragsurvey} integrates this retrieval chain into a complete pipeline: given a query, the system retrieves the most relevant documents from a knowledge base. This approach leverages large volumes of information without needing to retrain models, making it especially well-suited to domains with accumulated knowledge such as technical support.

\subsection{Technical support automation}

The application of natural language processing techniques to ticketing systems has been explored in various contexts. Potharaju et al.~\cite{potharaju2013tickets} analysed corpora of network trouble tickets to infer problems automatically, while Zhou et al.~\cite{zhou2016ticketrouting} proposed machine learning methods for intelligent ticket routing. These works focus on classification and triage, but do not address semantic retrieval over accumulated ticket histories.

In the HPC domain specifically, published efforts have focused predominantly on job monitoring and automatic diagnosis systems, whereas the systematic exploitation of knowledge contained in support histories has received less attention. \textsc{Fragata} occupies precisely this niche: high-quality semantic retrieval over a heterogeneous corpus dominated by real technical e-mail, with the operational requirements of a production service.

\section{Data sources and preparation}

\begin{figure*}[t]
\centering
\includegraphics[width=0.85\textwidth]{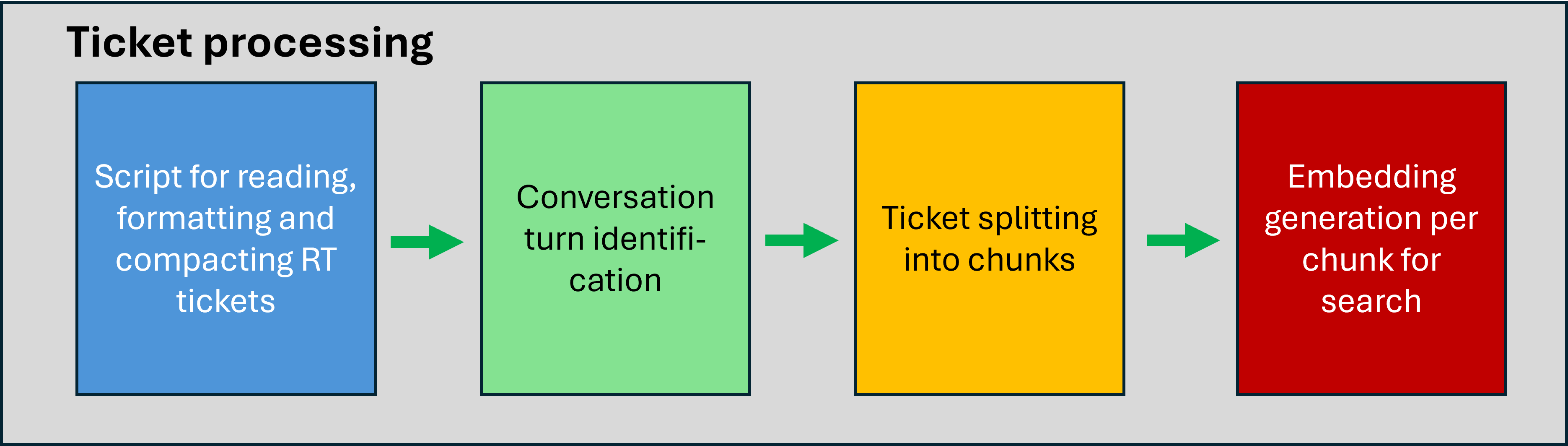}
\caption{Ticket processing pipeline: from SQL extraction of the RT history to the generation of \emph{embeddings} for semantic search.}
\label{fig:pipeline}
\end{figure*}

Figure~\ref{fig:pipeline} shows the overall data processing pipeline, which starts from the direct extraction of the RT history and culminates in the generation of vector representations (\emph{embeddings}) that enable semantic search.

\subsection{RT history dump}

The limitations of RT's search engine forced us to approach the problem from outside the tool. We developed specific SQL queries against RT~4.4.1's relational database that extract, for each ticket, the relevant metadata (identifier, queue, department, dates) and the entirety of the messages (\emph{transactions} and text-type \emph{attachments}) comprising the conversation thread between staff and users. The dump spans approximately twenty years of activity and is materialized as JSONL\footnote{JSONL (\emph{JSON Lines}) is a text format in which each line of the file constitutes an independent, valid JSON object. Unlike a conventional JSON file, JSONL allows line-by-line (\emph{streaming}) processing without loading the entire file into memory, making it especially suitable for incremental data pipelines.} files, a format chosen for its simplicity of streaming processing and its good fit with incremental pipelines.

\subsection{Normalization and chunking}
\label{sec:normalization}

As noted above, most interactions in RT occur via e-mail. When a user replies to a message, the e-mail client automatically includes the text of previous messages in the conversation thread. This causes the raw messages stored in each ticket to contain references to prior messages, leading to cumulative redundancy: the same information appears partially or fully repeated across multiple thread entries. Additionally, messages contain noise typical of e-mail: quoted headers, personal signatures, institutional banners, and inconsistent formatting.

The normalization phase removes this redundant information and associated noise by applying a sequence of operations: character decoding, removal of known headers and signatures, suppression of quoted messages from earlier in the thread, and whitespace collapsing. The goal is to present the retrieval system with each piece of information contained in a ticket exactly once, without duplications that could distort the results.

After normalization, a chunking process segments each conversation into overlapping fragments of bounded length. This segmentation is necessary because embedding models have a limit on the text length they can process, and shorter, more focused fragments produce more precise vector representations. Each chunk preserves identity metadata (\texttt{ticket\_id}, \texttt{conversation\_id}, \texttt{chunk\_id}), filtering metadata (\texttt{department}, \texttt{last\_updated}, \texttt{source\_type}), and ingestion traceability metadata.

\subsection{Complementary sources}

In addition to the ticket history, \textsc{Fragata} can ingest external technical documentation relevant to CESGA users: the centre's own web pages, scientific application manuals in PDF format, and repository documentation (READMEs and GitHub/GitLab wikis). Each source has a dedicated pipeline that produces rows in the same JSONL document, allowing them to be indexed jointly with the tickets and return unified results.

\section{Hybrid retrieval architecture}

The core of \textsc{Fragata} is a retrieval engine implemented in Python on FastAPI\footnote{FastAPI (\url{https://fastapi.tiangolo.com}) is a modern, high-performance Python framework for building web APIs, with native support for asynchronous operations and automatic documentation generation.}, instantiated from a declarative configuration (\texttt{config/rag.yaml}) and served through a lifecycle manager (\texttt{EngineManager}) described in Section~\ref{sec:hotswap}.

\subsection{Initialization and artifacts}

At startup, the engine detects the available compute device: CPU or GPU (via NVIDIA's CUDA platform). GPU usage significantly accelerates embedding computation and vector searches, reducing response and indexing times. The engine then loads the \texttt{paraphrase-multi\-lingual-MiniLM-L12-v2} model from the \texttt{sentence-trans\-form\-ers} family~\cite{reimers2019sentencebert}, which produces 384-dimensional embeddings with support for over 50 languages, making it especially well-suited to CESGA's trilingual corpus.

The engine reconstructs the FAISS~\cite{johnson2019faiss}\footnote{FAISS (\emph{Facebook AI Similarity Search}, \url{https://github.com/facebookresearch/faiss}) is an open-source library developed by Meta for efficient similarity search over large-scale dense vectors.} index from on-disk artifacts. FAISS can search millions of vectors to find those closest to a given query vector, which constitutes the basis of semantic retrieval: finding the text fragments whose meaning is most similar to the user's question. To avoid duplicating the corpus in memory, documents are reconstructed from FAISS's internal docstore rather than re-reading the complete JSONL.

In parallel, a BM25~\cite{robertson2009bm25}\footnote{BM25 (\emph{Best Matching 25}) is a classical information retrieval algorithm that scores documents based on the frequency of query terms, adjusted for document length. Unlike embeddings, it operates on exact word matches.} retriever is built over the same corpus, and the \texttt{mmarco-mMiniLMv2-L12-H384-v1}~\cite{bonifacio2022mmarco} reranker is loaded---a cross-encoder trained on mMARCO, the multilingual version of the MS~MARCO benchmark corpus. This model jointly evaluates the query and each candidate document to produce a relevance score more precise than the initial retrieval.

\subsection{Query variants}

Before retrieval, the engine generates weighted variants of the original query to maximize the probability of finding relevant documents. The variants include: the canonically normalized form of the query; a single-edit-distance correction when a frequent candidate is detected in the corpus vocabulary; an intent-based expansion that adds terms associated with the query type (definition, installation, containers, error/troubleshooting); and a translation via a curated Spanish/English term dictionary.

This last variant addresses a characteristic aspect of CESGA's environment: trilingualism. Support tickets contain text in Spanish, English, and Galician, as users freely employ all three languages in their communications. Translating key terms into English, the predominant language in technical documentation, standardizes query representation and improves matching against corpus fragments, regardless of the language in which they were originally written. Galician, due to its linguistic proximity to Spanish, implicitly benefits from both Spanish-language variants and the multilingual embeddings used by the system. This diversification increases recall on short or ambiguous queries and provides robustness against the linguistic heterogeneity of the corpus.

\subsection{Dense retrieval, lexical retrieval, and fusion}

For each variant, two channels are executed in parallel: a semantic search over FAISS (\texttt{semantic\_k} candidates) and a BM25 search (\texttt{lexical\_k}). The candidate pool sizes are dynamically adjusted when the query is very short, when a temporal filter is applied, or when results are restricted to a specific department, to compensate for subsequent pruning. The fusion of results from both channels employs \emph{Weighted Reciprocal Rank Fusion} (WRRF)~\cite{cormack2009rrf}:
\begin{equation}
  s(d) \;=\; \sum_{r \in R} w_r \cdot \frac{1}{k_{\text{rrf}} + \mathrm{rank}_r(d)},
\end{equation}
where $w_r$ combines the global channel weight (semantic or lexical) and the query variant weight. WRRF is robust against the different score scales of BM25 and FAISS and the heterogeneity introduced by the variants, since it operates solely on relative positions (rankings) rather than absolute scores.

\subsection{Query-aware reranking}

From the fused set, a number of candidates $\max(\text{rerank\_top\_n}, 4 \cdot \text{final\_k})$ is selected, optionally augmented with a semantic rescue and, for single-token queries, an exact lexical match rescue. The cross-encoder scores each candidate against several prompts (original, capitalization variants, derived variants) and the best score is retained.

Score adjustments (\emph{boosts} and penalties) based on domain heuristics are then applied. These adjustments modify each candidate's score based on specific signals: a document is boosted when the query explicitly suggests a department and the document belongs to that department; non-ticket sources are penalized in short queries where tickets are more likely to be relevant; low-information-density fragments receive a reduced score; titles covering all query terms are favoured; and exact matches are boosted for single-term queries. These adjustments incorporate domain knowledge from HPC technical support that generic neural models do not capture on their own.

\subsection{Ticket-level post-processing}

The engine operates internally at the chunk level, but the support staff is presented information at the ticket level via a link to the RT service hosted at CESGA. The API layer deduplicates by \texttt{ticket\_id}, applies adaptive overfetch when deduplication leaves the response short, and produces clean text snippets by removing redundant prefixes and incorporating page titles when they provide context. This separation between internal ranking (fragment) and external presentation (ticket) makes it possible to optimize retrieval quality without polluting the information presented to the user.

\section{Incremental ingestion and hot-swap}
\label{sec:hotswap}

A production search service cannot afford downtime for re-indexing. \textsc{Fragata} addresses this requirement with an ingestion architecture orchestrated by Slurm, the job scheduler of FinisTerrae~III, and an atomic engine reload mechanism.

\subsection{Job model}

Every ingestion, whether initiated by API (\texttt{/ingest/web}, \texttt{/ingest/pdf}, \texttt{/ingest/repo-docs}, \texttt{/ingest/rt-weekly}) or by the scheduler, materializes as a job with a manifest (\texttt{manifest.json}) and an execution log (\texttt{job.log}) persisted atomically. The manifest tracks state (\texttt{queued}, \texttt{running}, \texttt{succeeded}, \texttt{failed}), current stage, progress, validated request, and per-stage metrics.

\subsection{Serialized critical section}

The stages that mutate global state, dataset merge, source catalogue update, FAISS index append, and engine reload, execute under a global mutation lock. Without this serialization, two concurrent jobs could corrupt each other's dataset, append deltas over inconsistent snapshots, or reload the engine with artifacts from different generations. The accepted cost is reduced concurrency in the critical section, which is acceptable given the centre's usage pattern.

\subsection{Incremental append with atomic promotion}

The incremental append module loads the active FAISS index, adds the delta documents, validates index and docstore growth, writes to a staging location, re-validates, and only then atomically promotes the new index and rotates the previous one to a timestamped backup. In the online flow, any inconsistency triggers an explicit failure rather than silently launching a full rebuild, allowing immediate detection and diagnosis.

\subsection{Engine hot-swap}

After a successful index mutation, the orchestrator requests an engine reload. The manager uses two separate locks: one serializes builds and the other protects the pointer to the active engine. The new engine is built outside the final swap, and the changeover only takes effect if the build completes successfully, incrementing a generation counter. If the build fails, the service continues responding with the previous engine, ensuring graceful degradation without service interruption.

\subsection{Ingestion cycle and transactional watermark}

The weekly batch pipeline extracts an incremental window from RT determined by a reference timestamp (\emph{watermark}) plus a safety overlap. This overlap is necessary to ensure that tickets whose last modification occurred very close to the boundary of the previous window are not missed: clock differences between servers, in-flight transactions at the cutoff time, or data propagation delays could cause a ticket modified just before the watermark to go unrecorded. By slightly overlapping the windows, the system reprocesses a small number of already-known tickets, which are discarded by deduplication, but avoids coverage gaps.

The pipeline prepares data by department, consolidates a global delta, updates the index, and reloads the engine. The watermark is only confirmed if all previous stages have succeeded, making the temporal pointer advance a transactional operation with respect to the engine state.

\section{Deployment and FinisTerrae~III integration}

\begin{figure*}[t]
\centering
\includegraphics[width=0.85\textwidth]{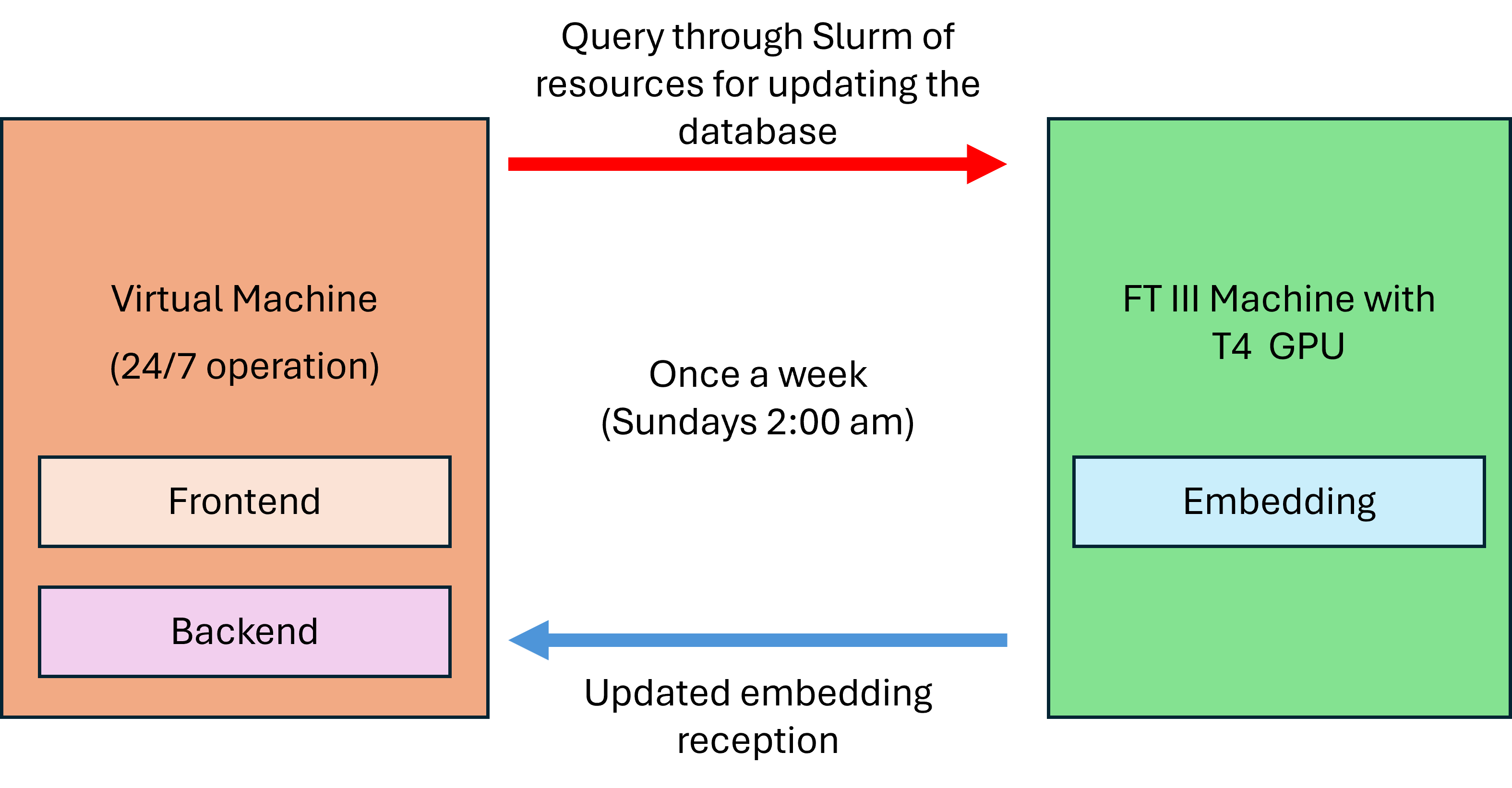}
\caption{Deployment topology: the virtual machine hosts the web service and the API, while expensive indexing stages are offloaded to the FinisTerrae~III supercomputer equipped with an NVIDIA T4 GPU.}
\label{fig:architecture}
\end{figure*}

\textsc{Fragata} is deployed in a virtual machine + supercomputer topology, illustrated in Figure~\ref{fig:architecture}. The virtual machine hosts the static frontend served by Nginx\footnote{Nginx (\url{https://nginx.org}) is a high-performance HTTP server and reverse proxy, widely used to serve static content and as an entry point that distributes requests among an application's internal services.} and the FastAPI backend managed by \texttt{systemd}. Communication between frontend and backend is configured via CORS (\emph{Cross-Origin Resource Sharing}), a browser security mechanism that controls which origins (domains) are allowed to make requests to a server different from the one that served the page. This configuration allows the frontend and backend to be separated into independent domains while maintaining service security.

The expensive ingestion and re-indexing stages can be offloaded to the FinisTerrae~III supercomputer (Figure~\ref{fig:architecture}) via an offload mechanism controlled by environment variables. The remote lifecycle materializes as observable stages (\texttt{resource\_requested}, \texttt{waiting\_resources}, \texttt{running\_remote}, \texttt{sync\_back}) whose metadata is persisted for traceability.

The resource release policy is configurable: in automatic mode, remote success implies implicit release; in explicit cancellation mode, job cancellation is also invoked on success; on workload failure, cancellation is always attempted. This integration makes it possible to re-index large volumes of the RT history without saturating the virtual machine, taking advantage of FinisTerrae~III nodes during periods of low occupancy.

\section{Preliminary results and discussion}

The system is currently deployed in CESGA's internal production environment. Qualitative results on real queries from the support team show that \textsc{Fragata} systematically retrieves old tickets that were invisible to RT's native search, especially in the following scenarios:

\begin{itemize}
\item Natural language queries in Spanish or Galician about incidents originally reported in English, thanks to dictionary-based translation and the multilingual embedding model.
\item Morphological variants or typos in scientific application names, which RT's strict lexical search could not resolve.
\item Intent-based queries (e.g., ``how to install X'' or ``error when running Y'') where query meaning matters more than the exact words used.
\item Combined searches by content and date range or department.
\end{itemize}

The main limitations identified are the dependence on the quality of upstream e-mail cleaning, intent heuristics based on static dictionaries rather than learned models, a curated ES/EN vocabulary instead of a general translation system, and a reranking cost that grows with the product of the number of candidates and the number of query variants.

\section{Conclusions and future work}

This paper has presented \textsc{Fragata}, a semantic retrieval system for HPC support tickets based on hybrid RAG and deployed on CESGA's infrastructure. The combination of dense retrieval via embeddings, BM25 lexical retrieval, WRRF-weighted fusion, and cross-encoder reranking, together with an incremental ingestion architecture with engine hot-swap and selective offloading to the FinisTerrae~III supercomputer, has made it possible to turn over twenty years of Request Tracker history into an effectively searchable resource, overcoming the limitations of RT~4.4.1's native search.

Future work includes the incorporation of an automatically learned intent classifier in place of static heuristics, quantitative evaluation with a curated query set and relevance judgments generated by the support team itself, extension of the system to a conversational interface with augmented generation that synthesizes actionable answers citing source tickets, and generalization of the deployment to other HPC centres interested in exploiting their own support histories.

\section*{Acknowledgements}

This research project was made possible through the access granted by the Galicia Supercomputing Center (CESGA) to its supercomputing infrastructure. The supercomputer FinisTerrae III and its permanent data storage system have been funded by the NextGeneration EU 2021 Recovery, Transformation and Resilience Plan, ICT2021-006904, and also from the Pluriregional Operational Programme of Spain 2014-2020 of the European Regional Development Fund (ERDF), ICTS-2019-02-CESGA-3, and from the State Programme for the Promotion of Scientific and Technical Research of Excellence of the State Plan for Scientific and Technical Research and Innovation 2013-2016 State subprogramme for scientific and technical infrastructures and equipment of ERDF, CESG15-DE-3114

Additionally, this work was carried out within the framework of the Technological Upgrade Project for the Computing and Data Node of the Galicia Supercomputing Center (CESGA), funded by the Recovery, Transformation and Resilience Plan through the NextGenerationEU instrument of the European Union, within the Strategic Project for Economic Recovery and Transformation in Microelectronics and Semiconductors (PERTE Chip), in accordance with Royal Decree 714/2024.

\bibliographystyle{Jornadas}
\bibliography{biblio}

\end{document}